\documentstyle{mn2e}
\input epsf

\def\plotone#1{\centering \leavevmode
\epsfxsize=\columnwidth \epsfbox{#1}}

\def\plotthree#1#2#3{\centering \leavevmode
\epsfxsize=.65\columnwidth \epsfbox{#1} \hfil
\epsfxsize=.65\columnwidth \epsfbox{#2} \hfil
\epsfxsize=.65\columnwidth \epsfbox{#3} \hfil}

\def\spose#1{\hbox to 0pt{#1\hss}}

\def\approxlt{{\mathrel{\spose{\lower 3pt\hbox{$\sim$}}
        \raise 2.0pt\hbox{$<$}}}}
\def\approxgt{\mathrel{\spose{\lower 3pt\hbox{$\sim$}}
        \raise 2.0pt\hbox{$>$}}}

\newcommand{\drv}[2]{\frac{\partial #1}{\partial #2}   }

 \title[Impact of stochastic gas motions on galaxy cluster abundance
profiles]{Impact of stochastic gas motions on galaxy cluster abundance
profiles}

\author[P.~Rebusco et al.]{P.~Rebusco $^{1}$, E.~Churazov$^{1,2}$, H.~B\"ohringer$^{3}$, W.~Forman$^{4}$ \\
$^1$ Max-Planck-Institut f\"ur Astrophysik, Karl-Schwarzschild-Strasse 1, 85741
Garching, Germany\\
$^2$ Space Research Institute (IKI), Profsoyuznaya 84/32, Moscow 117810, 
Russia\\
$^3$ MPI f\"{u}r Extraterrestrische Physik, P.O. Box 1603, 85740
Garching, Germany\\
$^4$ Harvard-Smithsonian Center for Astrophysics, 60 Garden St.,
Cambridge, MA 02138, USA
}

\pagerange{\pageref{firstpage}--\pageref{lastpage}}
\pubyear{2004}

\begin{document}

\maketitle
\label{firstpage}

\begin{abstract}
The impact of stochastic gas motions on the metal distribution in
cluster core is evaluated. Peaked abundance profiles are a
characteristic feature of clusters with cool cores and abundance
peaks are likely associated with the brightest cluster galaxies
(BCGs) which dwell in cluster cores. The width of the abundance peaks
is however significantly broader than the BCG light distribution,
suggesting that some gas motions are transporting metals originating
from within the BCG. Assuming that this process can be treated as
diffusive and using the brightest X-ray cluster A426 (Perseus) as an
example, we estimate that a diffusion coefficient of the order of
$2\times10^{29}~{\rm cm^2~s^{-1}}$ is needed to explain the width of the
observed abundance profiles. Much lower (higher) diffusion
coefficients would result in too peaked (too shallow) profiles. Such
diffusion could be produced by stochastic gas motions and our analysis
provides constraints on the product of their characteristic velocity
and their spatial coherence scale.  We speculate that the
activity of the supermassive black hole of the BCG is driving the
stochastic gas motions in cluster cores. When combined with the
assumption that the dissipation of the same motions is a key gas
heating mechanism, one can estimate both the velocity and the spatial
scale of such a diffusive processes.
\end{abstract}


\section{Introduction}
Heavy metals are observed through X-ray spectroscopy of the hot gas
in galaxy clusters through the emission lines of highly ionized Ca,
Si, S, Fe and other elements. These elements are produced in stars and
subsequently injected into the intracluster medium (ICM). On average,
the metallicity of the cluster ICM is $\sim 1/3$ of the solar
value and it does not seem to vary significantly at least up to a redshift of
$\sim$1 (e.g. Mushotzky \& Loewenstein 1997, Tozzi et al.
2003). This lack of evolution and the relative abundances of the
different elements suggest an early enrichment of the ICM by
type II supernovae (e.g. Finoguenov et al. 2002).  With the high
spatial and energy resolution of ASCA, Beppo-Sax, Chandra and
XMM-Newton, the radial distribution of metals has been recently mapped
for a large sample of nearby clusters (e.g. Fukazawa et al. 2000,
Irwin \& Bregman 2001, De Grandi \& Molendi 2001, Schmidt, Fabian \&
Sanders 2002, Matsushita et al. 2002, Churazov et al. 2003, De Grandi et
al. 2004).  Clusters can be divided into two groups, depending on
their X-ray properties (Jones \& Forman 1984): clusters with cool
cores (having a peaked surface brightness profile and a cool core
centered at the BCG) and clusters without cool cores (having more or
less a flat surface brightness profile in the core and no clear
evidence for a cool region near the center). These two groups are
traditionally called cooling flow and non-cooling flow clusters
(Fabian 1994). The spatial distribution of metals for these two
groups is also markedly different: clusters without cool cores have
a more or less uniform distribution, while clusters with cool
core have strongly peaked abundance profiles (sometimes exceeding
the solar abundance) centered at the BCG (e.g. Fukazawa et al. 1994,
2000, Matsumoto et al. 1996, De Grandi \& Molendi 2001).  Moreover
the relative abundances of different elements in the central peaks
of cool core clusters suggests that type Ia supernovae have
played a significant role in the enrichment process (e.g. Finoguenov
et al. 2002, see also Renzini et al. 1993). It is very likely
(e.g. B\"ohringer et al. 2004, De Grandi et al. 2004) that these
central abundance peaks are produced predominantly by the stars of the
brightest cluster galaxy after the cluster was assembled.  If so,
then the distribution of the metals should reflect the distribution
of the stars (i.e. light) of the BCG and the differences in the light
and metal distributions can be used as a proxy for processes which
transport and mix the metals of the ICM. Indeed in the best studied
cases such as M87, Centaurus or Perseus (Matsushita et al. 2003,
2004, Churazov et al. 2003) the metal distribution in the central
abundance peaks is broader than the BCG light distribution, implying
that injected metals diffuse to larger radii.

Another characteristic property of cool core clusters is a short gas
cooling time -- a factor of 10 to 100 shorter than the Hubble
time. Recent X-ray observations (e.g. Peterson et al. 2003, Matsushita
et al. 2002, Kaastra et al. 2004) however suggest that the gas does
not cool below temperatures of 1-3 keV and an external source of
energy is needed to compensate for the gas cooling losses. A number of
potential sources of energy have been considered - among them thermal
conduction (e.g. Narayan \& Medvedev 2001, Voigt \& Fabian 2004) or
gas motions of different origin. One of the latter family of models
utilizes outflows of relativistic plasma, driven by the central super
massive black hole (AGN), which interact with the ICM (e.g. Churazov
et al.  2001, 2002). The details of the outflow/gas interactions are
not yet understood completely and the assumptions on the character of
the induced gas motion vary from being almost pure radial (e.g. Fabian
et al. 2003) to mostly stochastic (e.g. Churazov et al. 2002). The
observed peaked abundance profiles provide a possibility to evaluate
the velocities and the spatial scales of these motions. The same
motions may affect the thermal balance of the gas in the core by
viscous dissipation or by tubulent transport of heat from larger radii
(Cho et al., 2003, Kim \& Narayan 2003, Voigt \& Fabian 2004). The
relative importance of both mechanisms is compared in Dennis \&
Chandran (2004).

In this paper we focus on the impact of stochastic gas motions of
arbitrary origin on the transport of metals from the BCG through
the ICM in cool core clusters.  We adopt here a simple
assumption that a diffusion approximation can crudely characterize
this transport process.  Using the brightest X-ray cluster, Perseus
(A426), as an example we estimate parameters of the stochastic motions
which are broadly consistent with the data.

The structure of the paper is the following.  In section
\ref{sec_model} the basic input data and the ingredients of the model
are described. In section 3 the evolution of the abundance profiles
is analyzed for various assumptions on the diffusion coefficient. The
results are discussed in section 4. The last section summarizes our
findings.

We adopt a Hubble constant of $H_0=70~km~s^{-1}~Mpc^{-1}$,
$\Omega_M=0.3$ and $\Omega_\Lambda=0.7$, which places the Perseus
cluster at 78 Mpc for a redshift of 0.018.

\section{The model}
\label{sec_model}

In this section the basic input parameters of the model are
described.

\subsection{A426 density, temperature and abundance profiles}
The electron density $n_e$ and the temperature $T_e$ profiles used here are
based on the deprojected XMM-Newton data (Churazov et al. 2003, 2004)
which are also in broad agreement with the ASCA (Allen \& Fabian
1998), Beppo-Sax (De Grandi \& Molendi 2001, 2002) and Chandra
(Schmidt et al. 2002, Sanders et al. 2004) data. Namely:
\begin{eqnarray}
n_e=\frac{4.6\times10^{-2}}{[1+(\frac{r}{57})^2]^{1.8}}+
\frac{4.8\times10^{-3}}{[1+(\frac{r}{200})^2]^{0.87}}~~~{\rm cm}^{-3}
\label{ne}
\end{eqnarray}
and
\begin{eqnarray}
T_e=7\times\frac{[1+(\frac{r}{71})^3]}{[2.3+(\frac{r}{71})^3]}~~~{\rm keV},
\label{te}
\end{eqnarray}
where $r$ is measured in kpc. The hydrogen number density is assumed
to be related to the electron number density as $n_{H}=n_{e}/1.2$.

\begin{figure}
\plotone{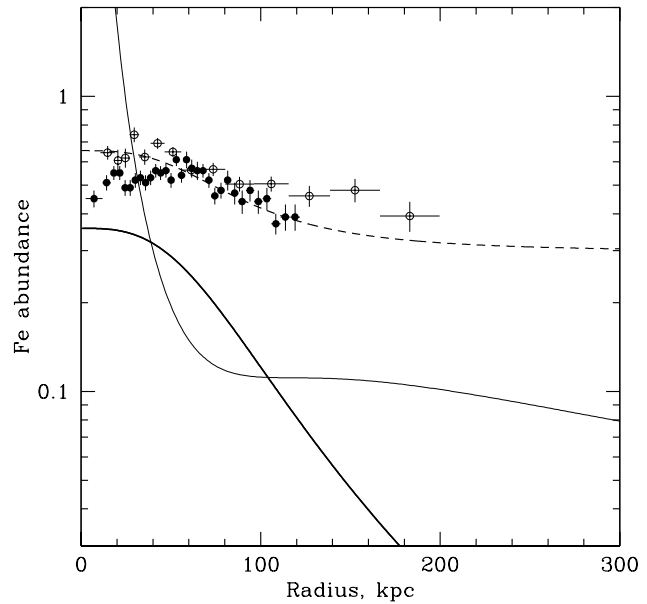}
\caption{Comparison of the observed and the expected iron abundance
profiles for the Perseus cluster. Solid circles correspond to the
Chandra measurements of projected abundance profile (from Schmidt et
al. 2002), while open circles correspond to the XMM-Newton deprojected profile
(from Churazov et al. 2003). The dashed line shows the abundance
profile adopted in this paper. This profile ignores the central
abundance decrement observed in Perseus and several other cooling flow
clusters. The thick solid line shows the same profile where a constant
value of $\sim$0.3 is subtracted. We assume below that this central
abundance excess (thick solid line) is primarily due to the metal
ejection of the central galaxy. For comparison we show the expected
iron abundance (thin solid line) due to the ejection of metals from
the galaxies, assuming a SNIa rate of 0.35 SNU, $k=1.4$ (see the text
for definitions) and a cluster age of 8 Gyr. The expected profile was
calculated assuming that the ejected metal distribution follows the
optical light.  The expected abundance profile is much more peaked
than the observed profile (due to the contribution of the central
galaxy) suggesting that some mechanism is needed to spread the metals.
}
\label{fig:prof}
\end{figure}

For the abundance  profile $a(r)$ (Fig.\ref{fig:prof}) we use the
deprojected profile obtained from XMM-Newton data (Churazov et
al. 2003). The normalization was reduced by $\sim$15\% percent to match
better the most recent deep Chandra observations of the Perseus core
(Schmidt et al. 2002, Sanders et al. 2004).  Namely:
\begin{eqnarray}
a(r)=0.3\times\frac{2.2+(r/80)^3}{~~1+(r/80)^3} a_\odot,
\label{eq:ab}
\end{eqnarray}
where $r$ is in kpc, and $a_\odot$ is the solar abundance (Anders \&
Grevesse 1989). The functional form used neglects completely the
central ``abundance hole'' observed in the Perseus cluster (Schmidt et
al. 2002, Churazov et al. 2003, Sanders et al. 2004) or in M87
(B\"ohringer et al. 2001, Matsushita et al. 2003). 
The nature of the abundance hole is unclear and it might be
related to ``visual'' effects (like the presence of a nonthermal
component in the X-ray continuum emission or resonant scattering, see
however Churazov et al. 2004, Gastaldello \& Molendi, 2004 for
arguments against resonant scattering) rather than to a real decrease
of the metal abundance in the very core. The solid line in
(Fig.\ref{fig:prof}) is the iron abundance profile from which a
constant value of $\sim$0.3 was subtracted. We assume below that the
central abundance excess is primarily due to the metal ejection from
the central galaxy. The total iron mass in the central excess is of
the order of $1.3\times10^9~M_\odot$ (or $\sim 5\times10^8~M_\odot$ within the
central 100 kpc). Compared to B\"ohringer et al. (2004) the lower
overall abundance normalization and the larger value of the subtracted
constant lead to somewhat lower masses attributed to the excess.

For comparison we show in Fig.\ref{fig:prof} the expected iron
abundance (thin solid line) due to the ejection of metals from the
galaxies, assuming a SNIa rate of 0.35 SNU, $k=1.4$ (see Section 2.3 for
definitions) and a cluster age of 8 Gyr. The expected profile was
calculated assuming that the ejected metal distribution follows the
optical light.  The expected abundance profile is much more peaked
than the observed profile (due to the contribution of the central
galaxy) suggesting that some mechanism is needed to spread the metals.

\subsection{The central galaxy: NGC 1275}
\begin{figure}
\plotone{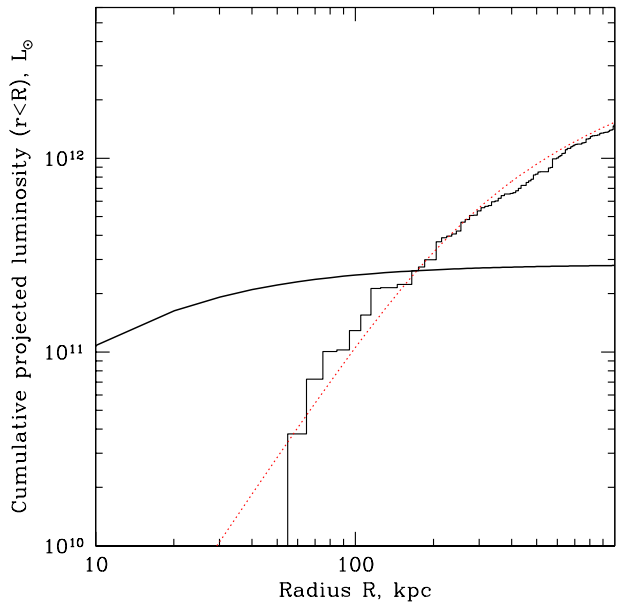}
\caption{Cumulative projected light distribution of the central galaxy
NGC1275 (thick solid line) and of  all the other galaxies in the cluster
excluding NGC1275 (histogram). The
positions and the magnitudes of the galaxies are taken from  Brunzendorf
\& Meusinger (1999). A simple King model fit to the light distribution
(omitting NGC1275) is shown. The central galaxy dominates light up to
a distance of $\sim$100 kpc.}
\label{fig:light}
\end{figure}

The light distribution of the central elliptical galaxy is described
here by a simple Hernquist profile (Hernquist 1990).  The effective
radius $r_e=15.3$ kpc and the total blue luminosity
$2.8\times10^{11}~L\odot$ are taken from Schombert (1987, 1988). While
the actual light profile of NGC1275 is more complicated than a single
Hernquist profile, this is an acceptable approximation for the purpose
of this study. In Fig.\ref{fig:light} the light distribution of the
central galaxy is compared with the light distribution due to all the
other cluster galaxies excluding NGC1275. The central galaxy
dominates up to a distance of $\sim$ 100 kpc and in subsequent
calculations we assume (unless explicitly stated otherwise) that 
the central excess of metals is produced by the central galaxy alone.

\subsection{Iron and gas injection rates}
\label{subsec_source}
The bulk of the cluster gas is enriched at early times of cluster
formation by a large number of type II supernovae. The central
abundance excess is believed to be formed at later times and the main
contributors are likely to be the type Ia supernovae and the stellar
mass loss associated with the central galaxy (Matsushita et al. 2003,
B\"ohringer et al. 2004, De Grandi et al. 2004). Both channels inject
 gas enriched with heavy elements, in particular - iron, into the
ICM. We use here  similar rates of iron injection by SNIa (eq. 4)
and stellar mass loss (eq. 5)  as in B\"ohringer et
al. (2004):
\begin{eqnarray}
\left (\frac{dM_{Fe}}{dt} \right )_{SN Ia} &  =  & SR
\times10^{-12} \left (\frac{L_B}{L_{\odot}^B} \right ) \eta_{Fe} \\ & = & R(t)~ 0.105\times 10^{-12}
\left ( \frac{L_B}{L_{\odot}^B} \right ) M_{\odot}~yr^{-1},\nonumber \\
\left ( \frac{dM_{Fe}}{dt} \right )_{*}& = &\gamma_{Fe}\times 2.5 \times
10^{-11} \left (\frac{t}{t_{H}} \right )^{-1.3} \\ \nonumber & & \times \left (
\frac{L_B}{L_{\odot}^B} \right ) M_{\odot}~yr^{-1},
\label{eq:rate}
\end{eqnarray}
 where $SR$ is the present SNIa rate in SNU (1 SNU - rate of
supernova explosions, corresponding to one SN event per century per galaxy with a blue luminosity of $10^{10}~L_{\odot}^B$), $\eta_{Fe}=0.7
M_{\odot}$ is the iron yield per SNIa, $\gamma_{Fe}=2.8\times10^{-3}$ is
the mean iron mass fraction in the stellar winds of an evolved stellar
population, $t_H$ is the Hubble time. The expression for the
stellar mass loss was adopted from Ciotti et al. (1991), assuming a
galactic age of 10 Gyr.  Along with heavy elements (iron) the stellar mass
loss can make a substantial contribution of hydrogen to the ICM. This
effect is important only in the central $\sim 20$ kpc and
it was neglected in the subsequent calculations. We therefore assume
that only heavy elements are injected to ICM by SNIa and stellar mass loss.
The time dependent
factor $R(t)= (t/t_{H})^{(-k)}$ takes into account an increased
SNIa rate in the past (Renzini et al. 1993), with the index $k$
ranging from 1.1 up to 2. 
A fiducial value for the present day 
SNIa rate of $0.15$ 
SNU (Cappellaro, Evans, \& Turatto 1999) was assumed. Thus the total
iron injection rate within a given radius $r$ can be written as
\begin{eqnarray}
s(<r,t)=\left ( \frac{dM_{Fe}}{dt} \right )_{*}+\left (
\frac{dM_{Fe}}{dt} \right )_{SN} \propto \left (
\frac{L_B(<r)}{L_{\odot}^B} \right )
\label{eq:source}
\end{eqnarray}

\begin{figure}
\plotone{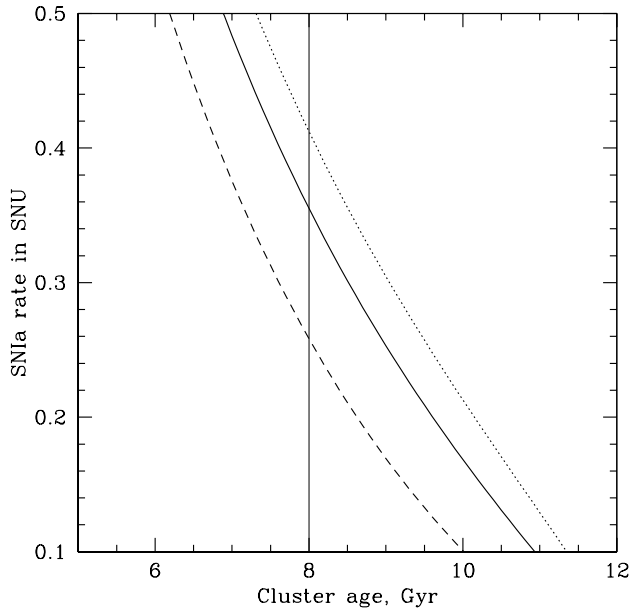}
\caption{SNIa rate in SNU required to produce the excess amount of
iron of $\sim 1.3\times10^9~M_\odot$ by the BCG with
$L_B=2.8\times10^{11}~L_\odot$ according to eq.\ref{eq:rate}. The solid line is
for $k=1.4$, the dotted one for $k=1.1$, the dashed one for $k=2$}
\label{fig:srage}
\end{figure}

The total amount of iron produced by the central galaxy during its
evolution from some initial time to the present is simply an
integral of equation (\ref{eq:source}) over time. The total amount of
iron in the central excess is set above (by our definition of the
central excess, see Fig.\ref{fig:prof}) to $\sim
1.3\times10^9~M_\odot$. Therefore, the parameters of the model ($SR$, $k$,
$t_{age}$) are constrained by the observed total iron mass (see
Fig.\ref{fig:srage}).

For the subsequent analysis we have selected two combinations of these
parameters: (0.35, 1.4, 8 Gyr) and (0.26, 2, 8 Gyr).  With these choices of
parameters one can readily predict the amount and the distribution of
the metals ejected from the galaxies into the ICM as displayed in
Fig.\ref{fig:prof}. 
In this figure we show the observed iron abundance and the expected iron
abundance, determined by comparing the amount of injected iron (by all
the galaxies, including the BCG) with the present day gas density
distribution. On large scales $\sim$30\% of the observed amount of
iron is provided by SNIa and stellar mass loss. The remaining
$\sim$70\% could be attributed to early enrichment by SNII. The
central excess of iron generated by the central galaxy is much more
peaked than the observed abundance profile. This is of course an
expected result, given that the effective radius of the galaxy is $\sim$15
kpc, while the characteristic size of the abundance excess is much
larger. The difference between the distribution of the ejected iron and
the observed abundance profile suggests that some process is transporting
the metals to larger radii. We assume below that this process can be
treated in a diffusion approximation.

\subsection{Diffusion of metals due to stochastic gas motions}
We assume that the ICM at a given radius $r$ is involved in stochastic
motions with a characteristic size sufficiently smaller than $r$ and
that this motion mixes the ICM.  We assume moreover that such motion
with a characteristic velocity $v$ and a characteristic coherence
length $l$ will lead to an effective diffusion coefficient of the
order of $D\sim\frac{vl}{3}$.  We further make the strong
simplification that the gas density and temperature do not evolve with
time, i.e. the gas entropy losses (due to radiation) and gains (due to
the dissipation of the stochastic motions and to the mixing of the low
entropy gas with the outer layers of higher entropy gas) cancel each
other (e.g. Churazov et al., 2002). In this approximation the
stochastic motions have a clear impact on the distribution of metals
and this process can be considered in a diffusion approximation:
\begin{equation}
\drv{n a}{t}=\nabla \cdot (D n \nabla { a} )+ S,
\label{eq:dif}
\end{equation}
where $n=n(r)$ is the gas density, $a=a(r,t)$ is the iron abundance
and $S=S(r,t)$ is the source term due to the iron injection from the
central galaxy, $D$ is the diffusion coefficient. Once the diffusion
coefficient is specified, eq.\ref{eq:dif} can be readily integrated.

\begin{figure}
\plotone{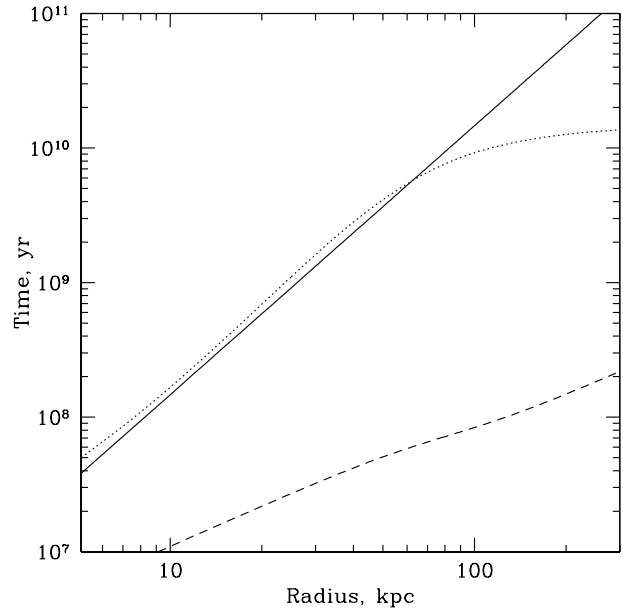}
\caption{Diffusion time scale (solid line) estimated as $t_{\rm diff}\sim
r^2/D$, where $D=2\times10^{29}~{\rm cm^2~s^{-1}}$. For comparison the
cumulative enrichment time (ratio of the amount of metals in the abundance
excess to the metal ejection rate by the central
galaxy ) is shown by the dotted line. The dashed line shows the sound
crossing time.}
\label{fig:tscale}
\end{figure}

A rough estimate of the required diffusion coefficient can be obtained
by comparing the characteristic time scales for diffusion and
enrichment (Fig.\ref{fig:tscale}). In this figure  the cumulative
iron enrichment time - i.e. the ratio of the observed amount of metals
in the abundance excess within a given radius $r$ is compared to
the current injection rate of iron by the central galaxy within the
same radius. The diffusion time was estimated as $t_{\rm diff}\sim r^2/D$,
where the diffusion coefficient $D$ is set to $2\times10^{29}~{\rm
cm^2~s^{-1}}$.  Fig.\ref{fig:tscale} suggests that diffusion with
coefficient $D\sim 2\times10^{29}~{\rm cm^2~s^{-1}}$ would significantly
affect the metal distribution in the core.

\section{Results}
\subsection{Constant diffusion coefficient}
\begin{figure*}
\plotthree{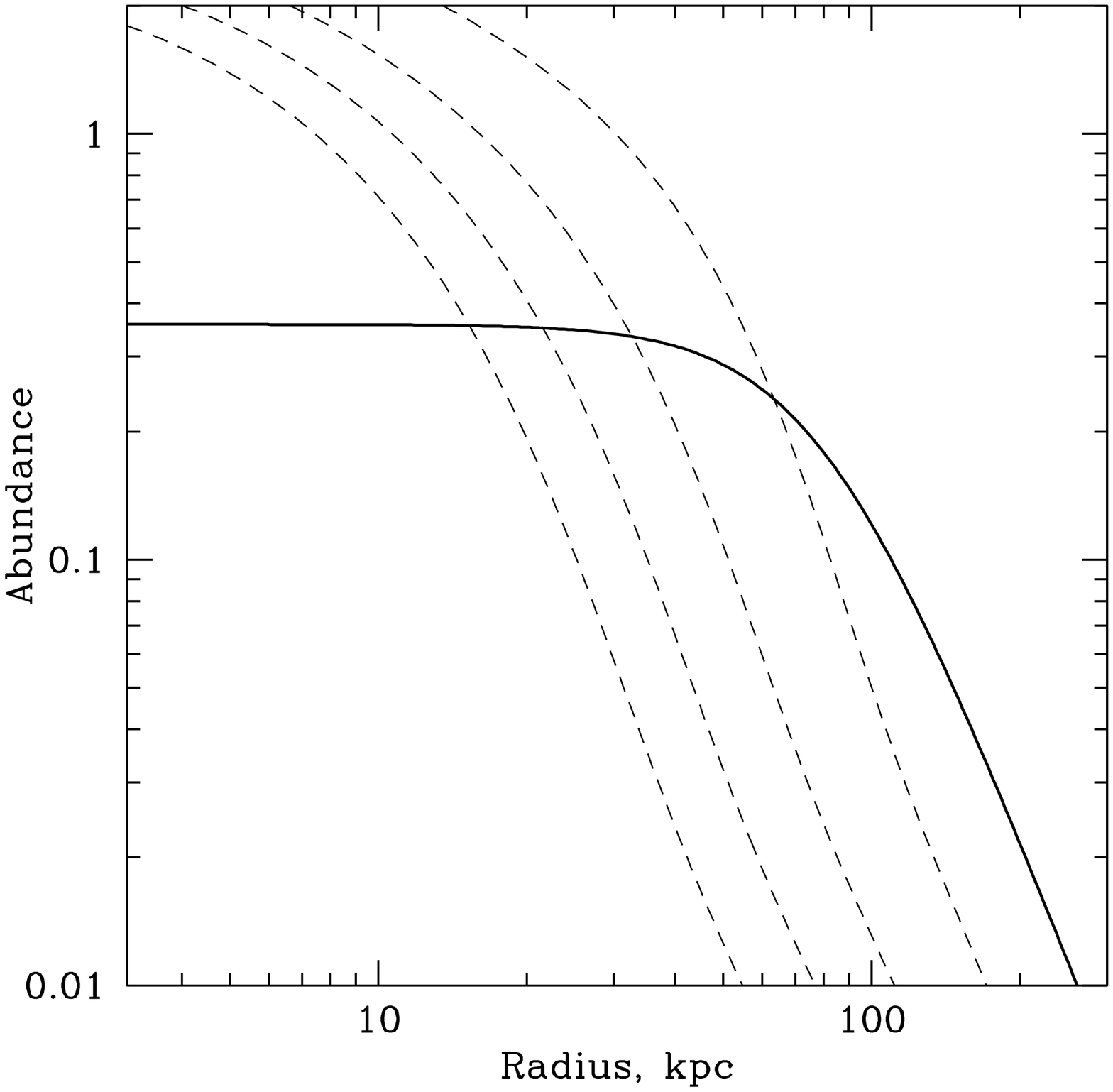}{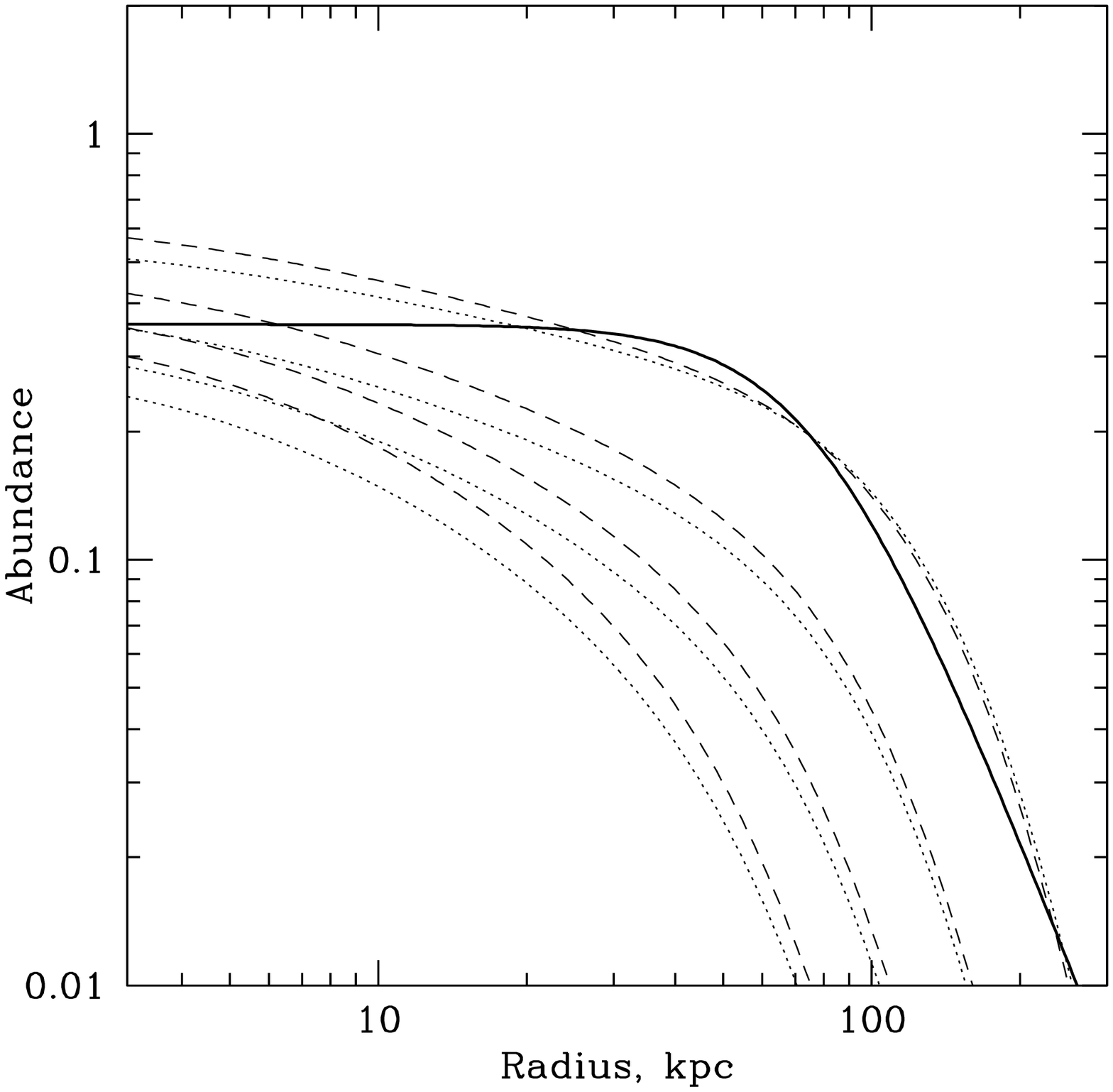}{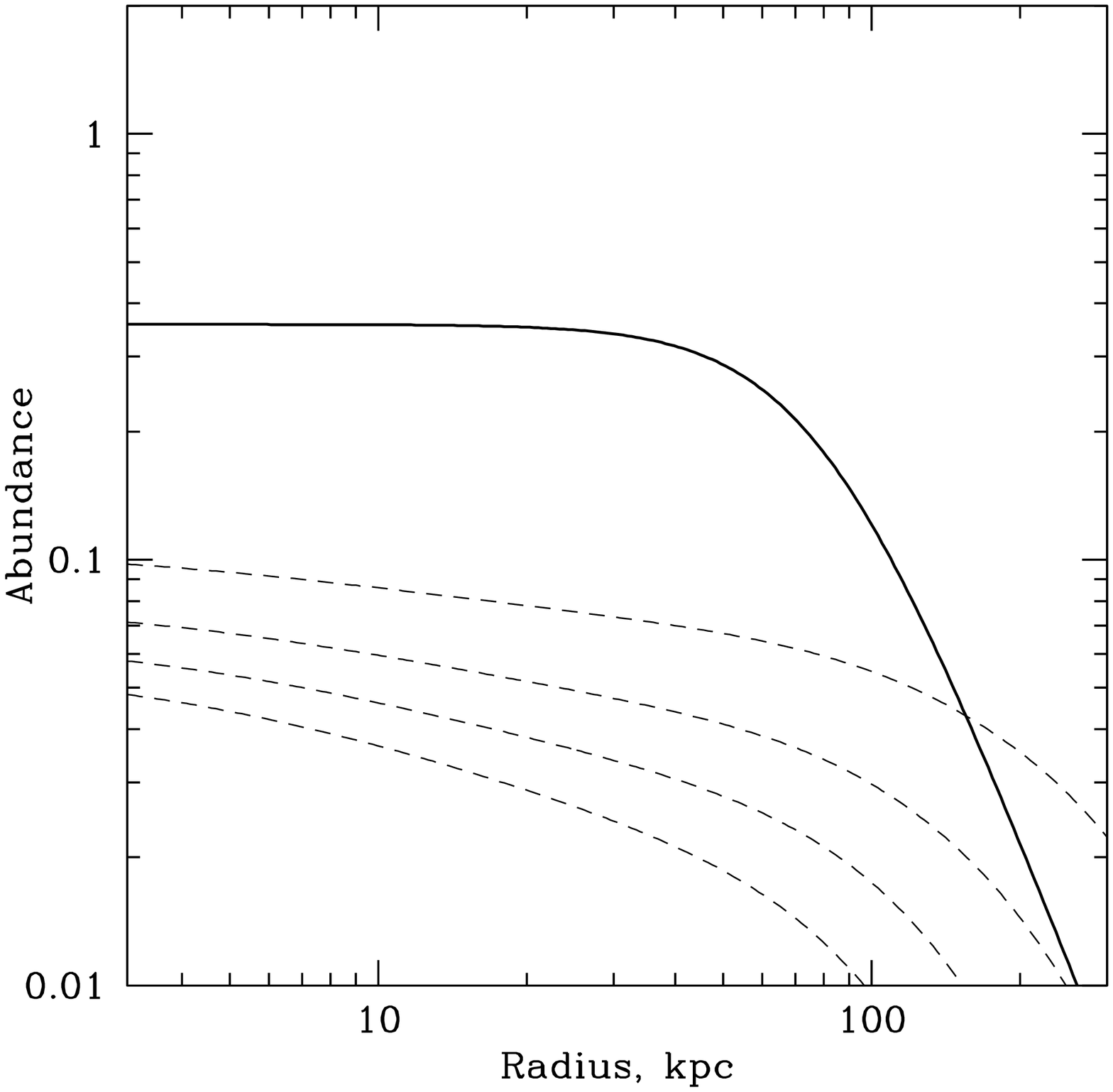}
\caption{Time evolution of the abundance profiles for different
diffusion coefficients $D=2\times10^{28}, 2\times10^{29}, 2\times10^{30} ~{\rm cm^2~s^{-1}}$
for the left, middle and right plots respectively. The  thick solid line is
the observed abundance peak, while the dashed lines show the abundance of
metals produced by the central galaxy during 1,2,4 and 8 Gyr (from the
bottom to the top)in the model with $k=1.4$ and $SR=0.35$. For
comparison in the middle plot the dotted line shows the same evolution for
$k=2$ and $SR=0.26$.}
\label{fig:m1m2}
\end{figure*}

In a more detailed modeling of the diffusion transport of 
metals in the ICM, we first consider the simplest case of a diffusion
coefficient which is independent of both time and radius.  
Fig.\ref{fig:m1m2} shows the evolution of the abundance peak
produced by the central galaxy after 1, 2, 4 and 8 Gyr for different
values of the diffusion coefficient: $D=2\times10^{28}, 2\times10^{29},
2\times10^{30} ~{\rm cm^2~s^{-1}}$ for the left, middle and right plots
respectively. The thick line in all plots shows the observed abundance
peak. It is clear that too small (or too large) diffusion coefficients
produce too peaked (or too shallow) abundance profiles, while
$D=2\times10^{29}~{\rm cm^2~s^{-1}}$ is roughly consistent with the
data. These calculations were done assuming $k=1.4$ and $SR=0.35$. The
comparison with the alternative model with $k=2$ and $SR=0.26$ (dotted
curves in the middle plot) shows that the differences in the shape of
the final abundance excess are minor.

\begin{figure}
\plotone{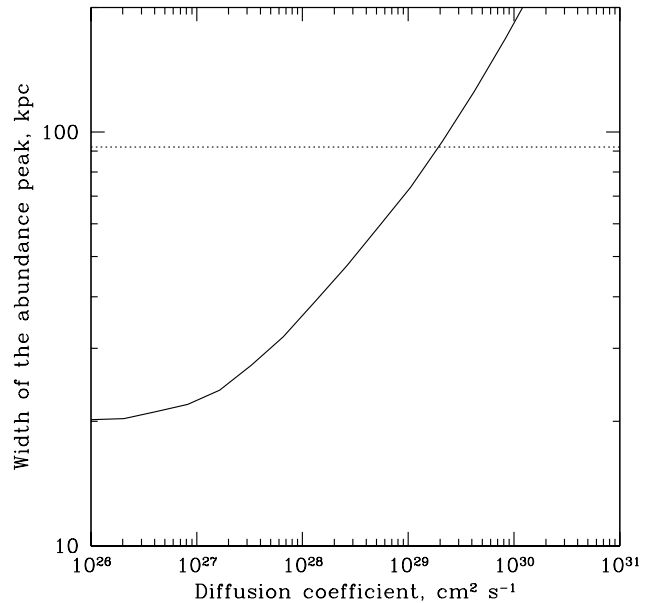}
\caption{Width of the abundance peak distribution (radius containing
half of the ejected iron) as a function of the diffusion coefficient
for $k=1.4$ and 8 Gyr cluster age. The dotted horizontal line shows
the width of the observed abundance excess.}
\label{fig:half}
\end{figure}

\begin{figure*}
\plotthree{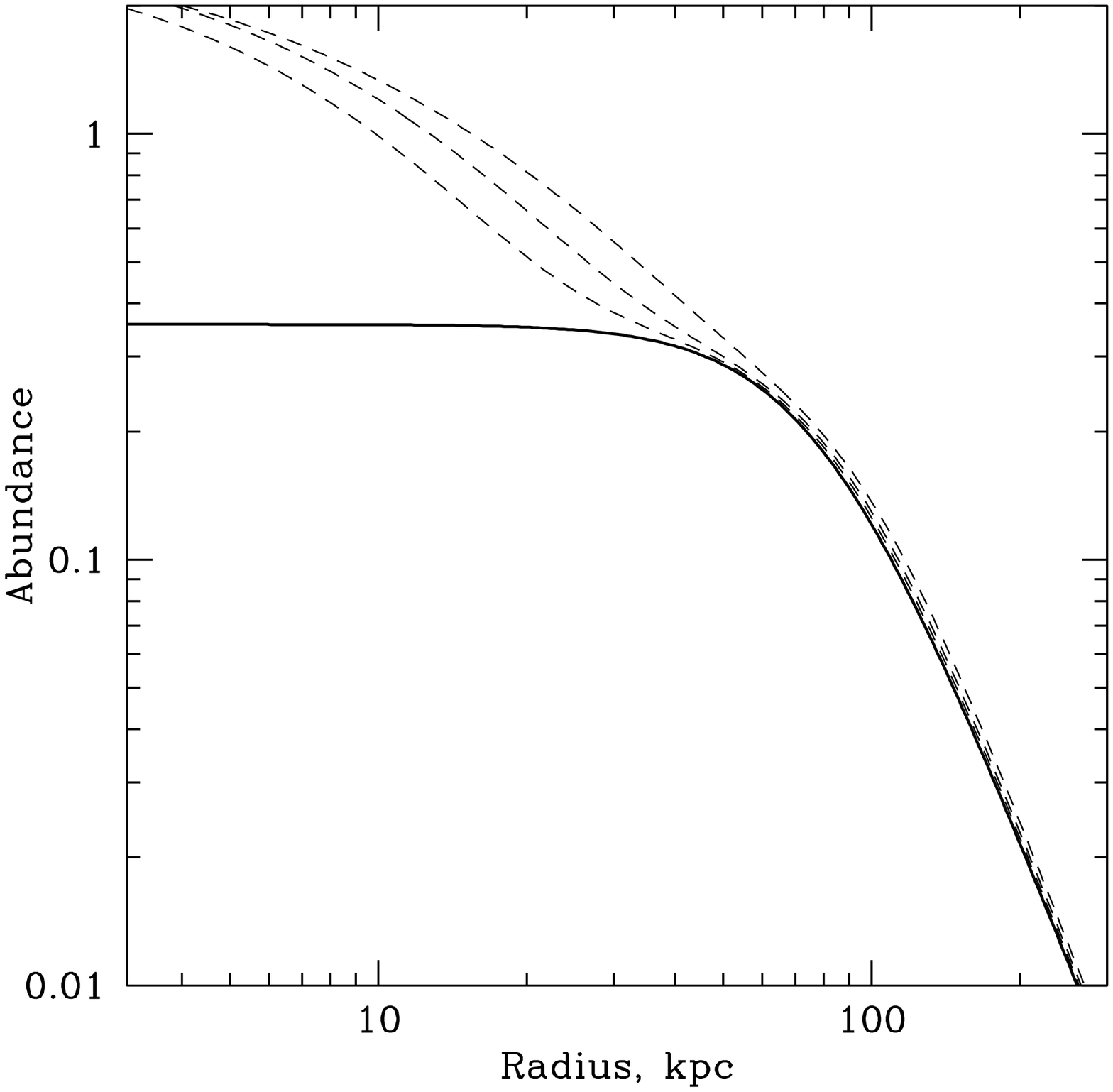}{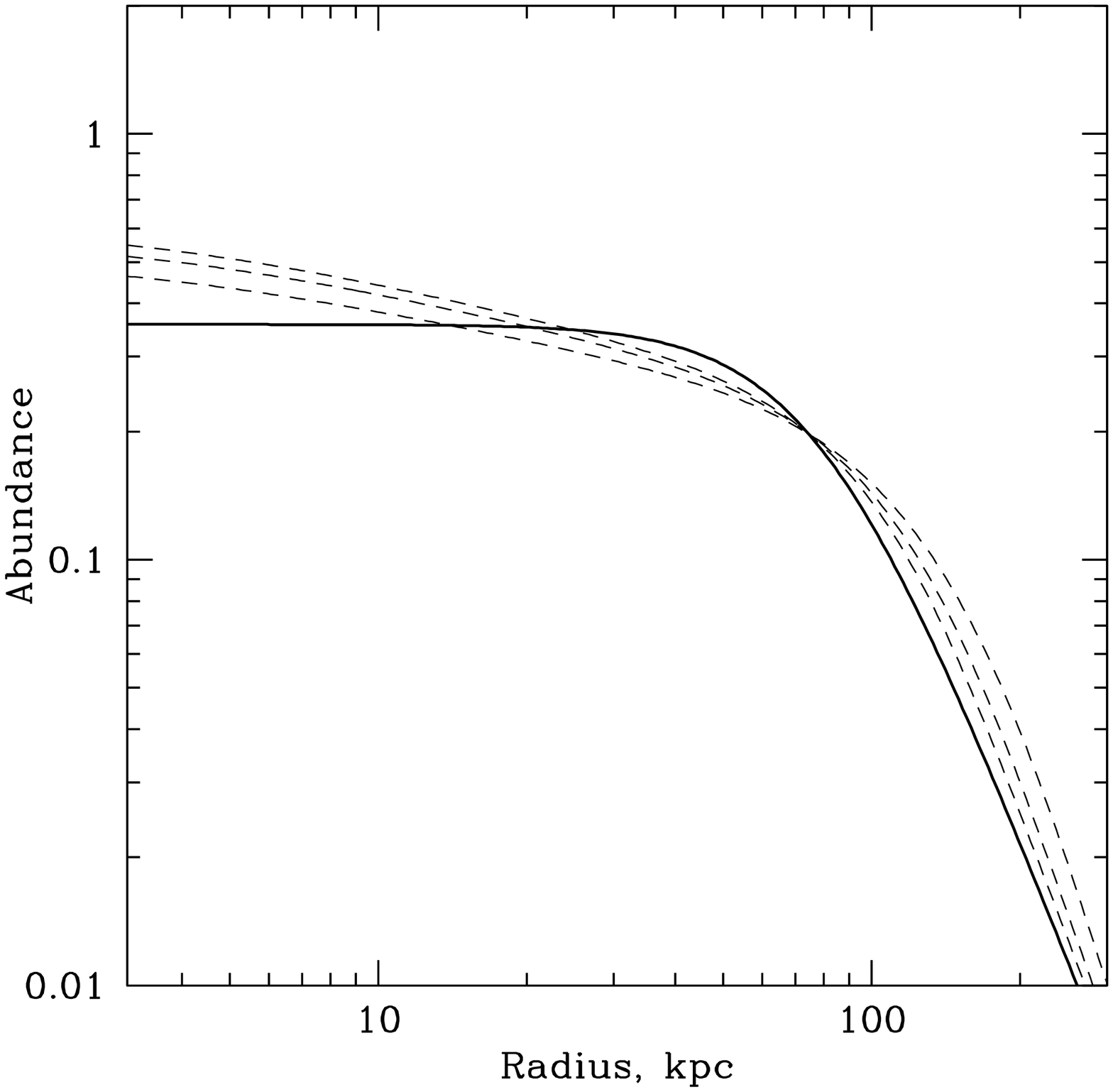}{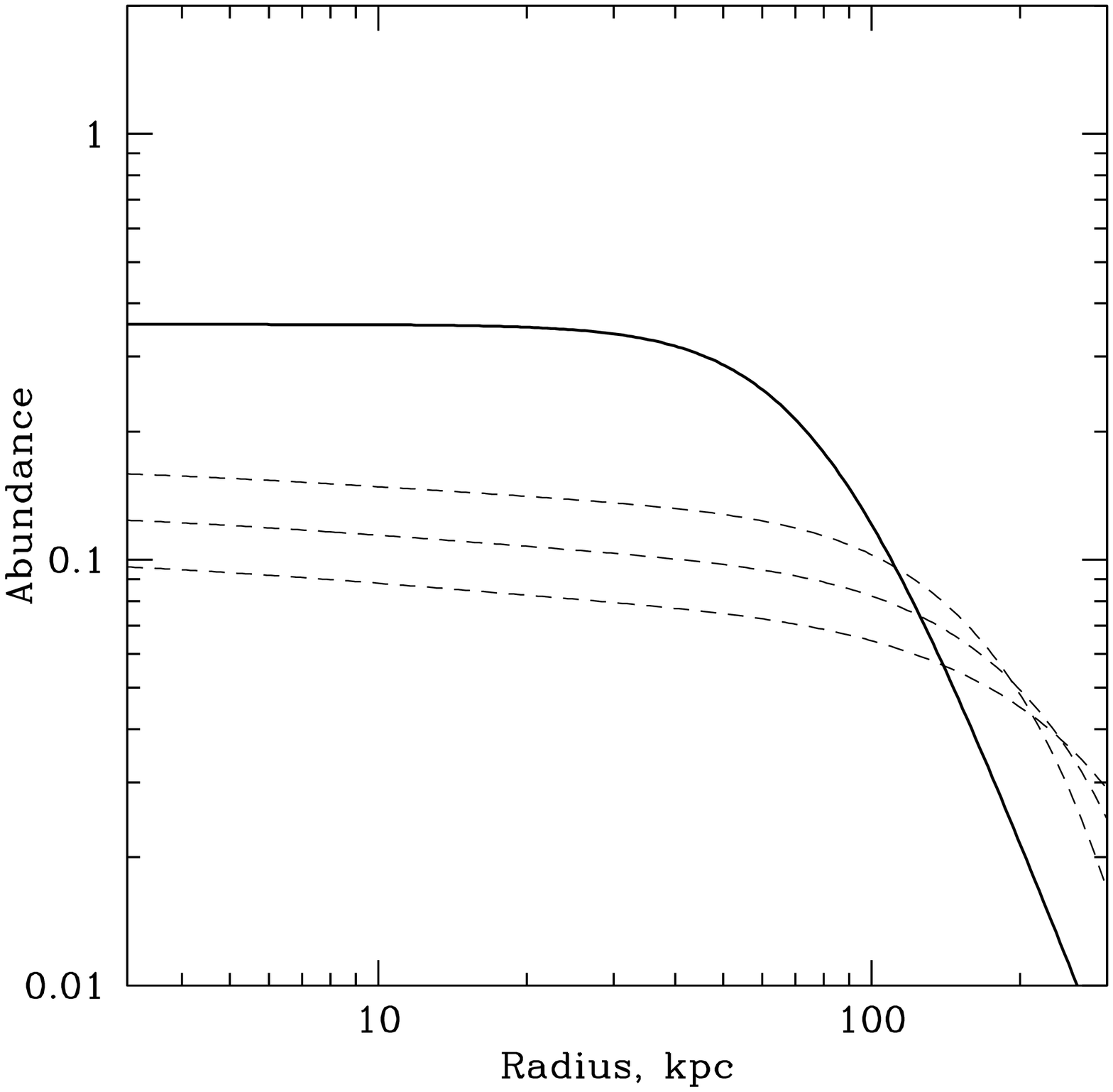}
\caption{Time evolution of the abundance profiles starting from the
observed abundance peak for different diffusion coefficients:
$D=2\times10^{28}, 2\times10^{29}, 2\times10^{30} ~{\rm cm^2~s^{-1}}$ for the left,
middle and right plots respectively. The initial profile is the observed
abundance peak distribution and the evolution of the profile is shown
after 1,2 and 4 Gyr (dashed curves from the bottom to the top) for the model
with $k=1.4$ and $SR=0.36$. The thick solid line is the observed abundance
peak. The calculations were done assuming metal injection rates
according to eq.\ref{eq:source}, starting from $t=t_H$. Note that for
the figures on the left(right) the profile quickly evolves to a peaked (shallow)
distribution, while for the central plot ($D= 2\times10^{29} ~{\rm
cm^2~s^{-1}}$) the evolution is very slow, approximately corresponding
to a quasi-steady state.}
\label{fig:now}
\end{figure*}

A similar characteristic value of the diffusion coefficient can be
derived by comparing the characteristic size of the observed
abundance excess and the excesses produced by  ejection and 
diffusion in the model. Fig.\ref{fig:half} shows the effective
radius (the radius containing half of the ejected metals) as a
function of the diffusion coefficient. For small diffusion
coefficients the metal distribution is essentially set by the light
distribution of the central galaxy. As the diffusion coefficient
increases the effective radius also increases. This plot also suggests
that $D\sim 2\times10^{29}~{\rm cm^2~s^{-1}}$ would provide a width of
the metal distribution similar to the one observed in the abundance
excess.
One can further consider the present day evolution of the abundance
profile, i.e. using the observed abundance peak as an initial condition,
as shown in Fig.\ref{fig:now}. Again, too small (or too large)
diffusion coefficients cause a quick steepening or flattening of the
abundance peaks on time scales of Gyr. The evolution of the profile for
the diffusion coefficient of the order of $2~10^{29}~{\rm
cm^2~s^{-1}}$ is much more gradual.

\subsection{Varying diffusion coefficient as a function of radius}
\begin{figure*}
\plotthree{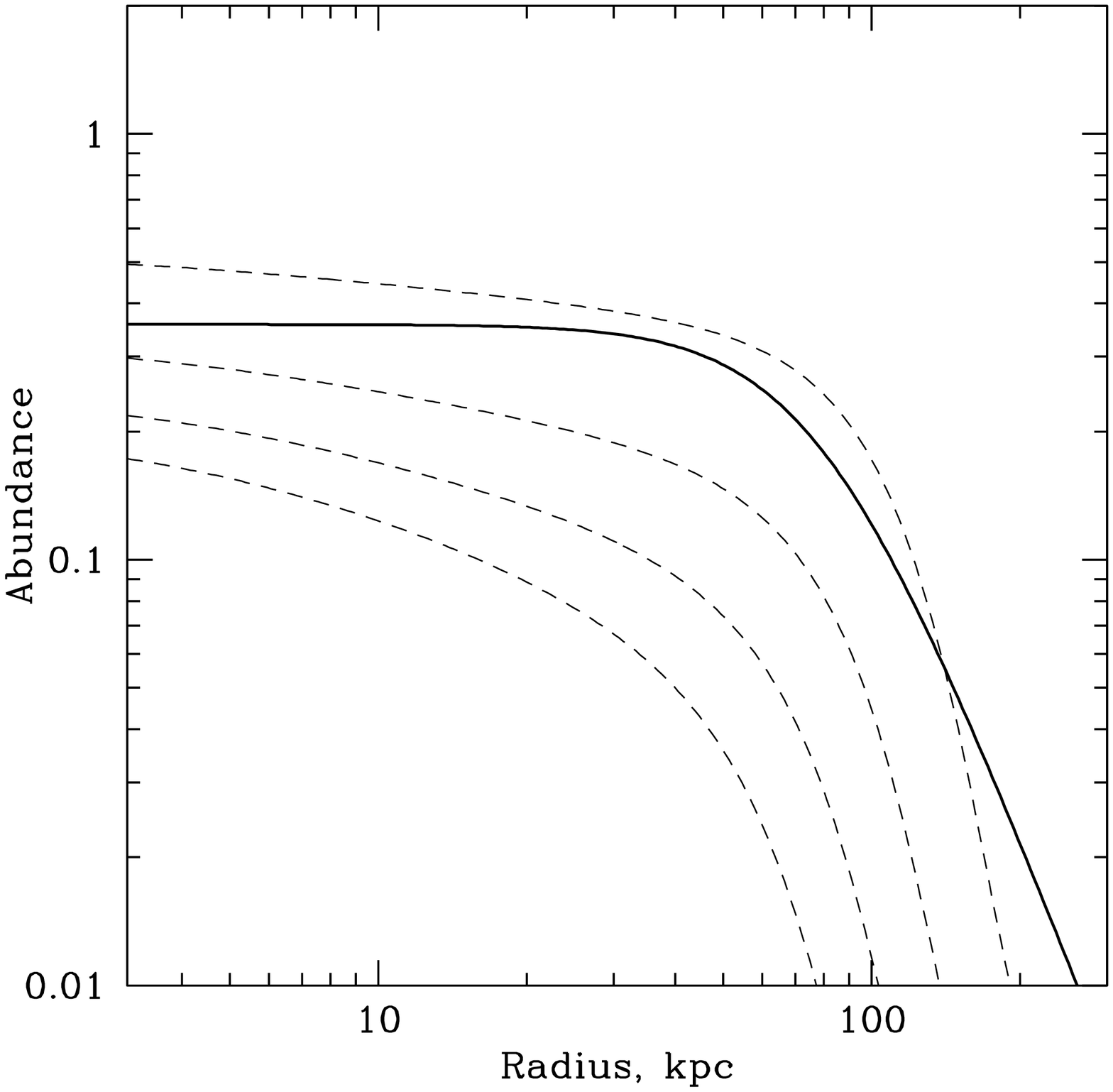}{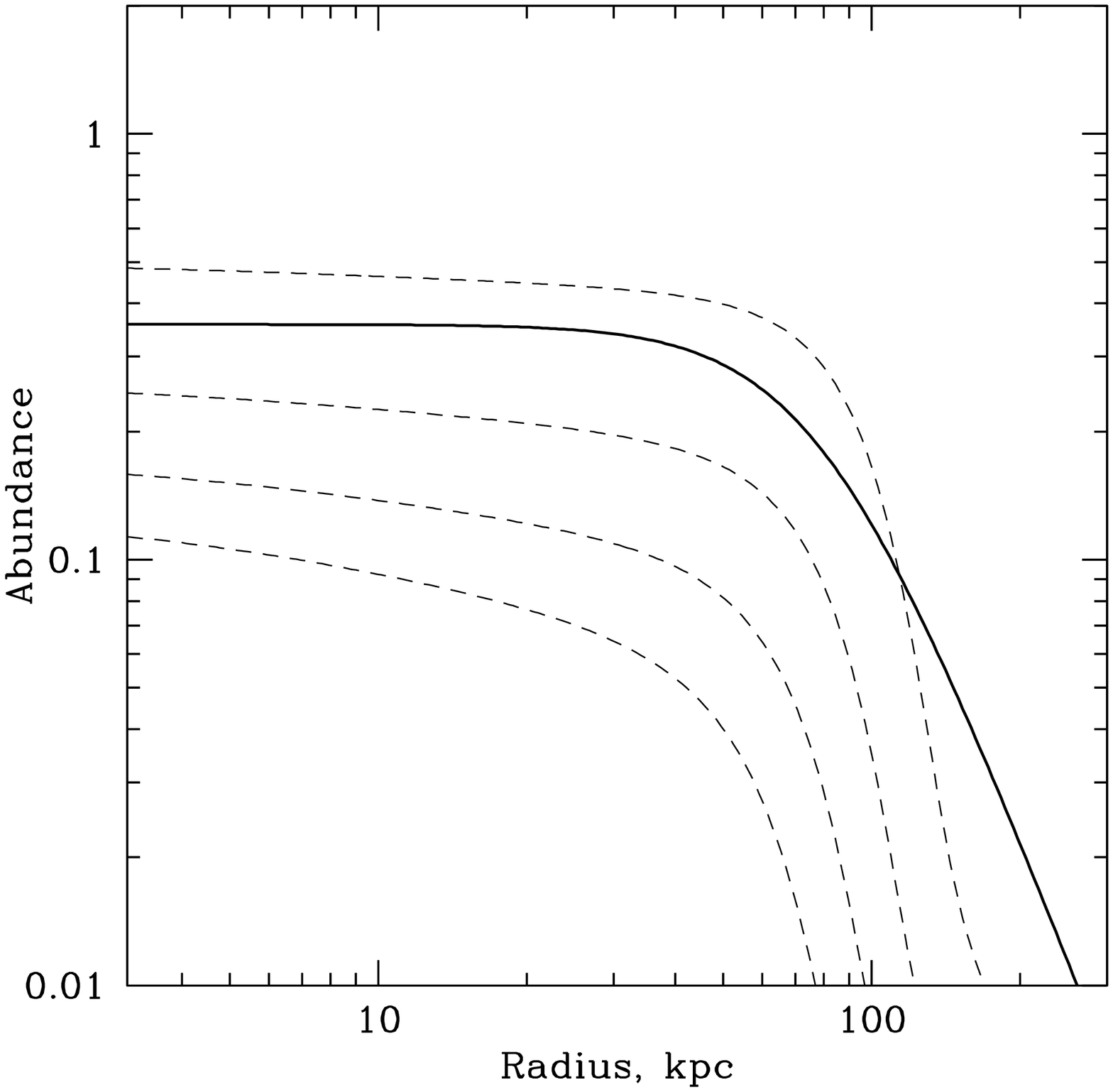}{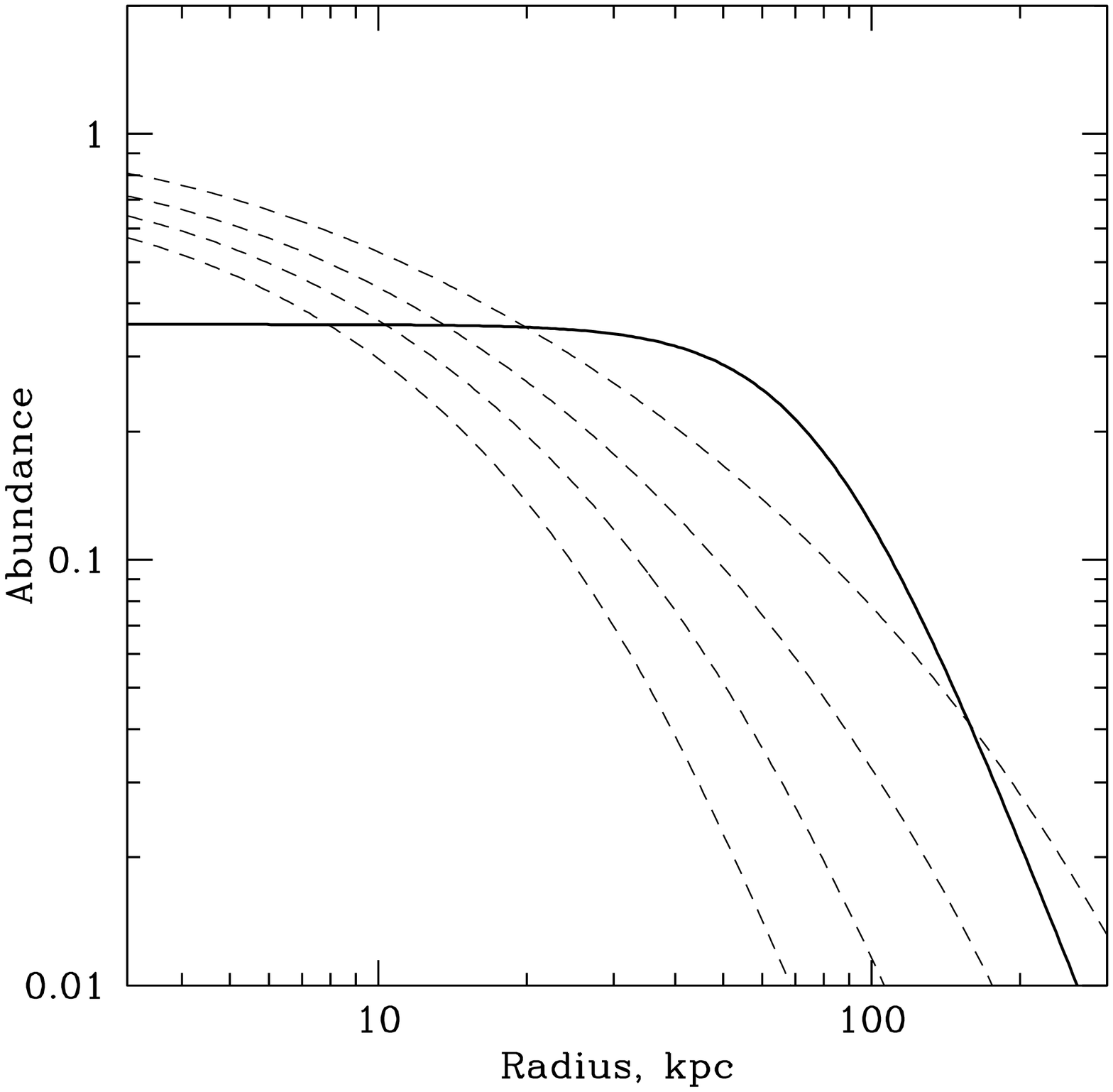}
\caption{Time evolution of the abundance profiles for different
radial dependence of the diffusion coefficient parametrized as $D=D_0
\left( n(r)/n(r_0) \right )^\alpha$. $\alpha=1,2,-1$ for the left, 
middle and right plots respectively. $D_0=2\times10^{29}~{\rm
cm^2~s^{-1}}$, $n$ is the gas density, $r_0=50$ kpc. The thick solid
line is the observed abundance peak, while the dashed lines show the
abundance of metals produced by the central galaxy during 1,2,4 and 8
Gyr ( in the model with $k=1.4$ and $SR=0.35$). As $\alpha$ increases
the distribution becomes more boxy, while for negative $\alpha$ the
distribution is more steep than the observed one.}
\label{fig:dindex}
\end{figure*}

In this section we estimate the impact of diffusion on the abundance
peaks when the diffusion coefficient is a function of the radius.  One
natural motivation for such an assumption is the hypothesis that an
AGN at the center of the BCG is driving the gas motions in the core.
In this case one would expect these motions to fade away outside the
cluster core, when the energy injected by the AGN into the ICM spreads
over large masses of gas.  Let us assume that a fraction $f$ of the
ICM volume at a given radius is involved in the stochastic motions
with characteristic velocity $v$ and  spatial scale $l$.  If
the dissipation of the stochastic motions with the rate
$\Gamma_{diss}\propto n v^3/l$ is the main source of energy which
offsets the gas radiative losses $\Gamma_{cool}\propto
n^2\Lambda(T)$, then $f\Gamma_{diss}\sim \Gamma_{cool}$ and
therefore $f\propto n$ (if $l$ and $v$ do not vary strongly with the
radius). The effective diffusion coefficient can then be written as
$D\propto n$, i.e. $D$ declines with the radius along with the
density. This is of course an oversimplified scaling, but to see the
influence of the diffusion coefficient changing with radius on the
abundance profile, we parametrized it as $D=D_0 \left( n(r)/n(r_0)
\right )^\alpha$, where $D_0=2\times10^{29}~{\rm cm^2~s^{-1}}$, $r_0=50$
kpc. The resulting profiles for $\alpha=1,2,-1$ are shown in
Fig.\ref{fig:dindex}. For declining diffusion coefficients, for
larger values of $\alpha$, the overall shape of the peak becomes more
boxy than for a constant diffusion coefficient (see
Fig.\ref{fig:m1m2} middle panel). In fact the profile with $\alpha=1$
is closer to the observed abundance peak than that with
a constant diffusion coefficient. However, given the uncertainties of
the adopted abundance peak and the oversimplified nature of the model,
one cannot claim that a variable diffusion coefficient provides a
significantly better description of the observed abundance peak.

The right plot in Fig.\ref{fig:dindex} is calculated for $\alpha=-1$,
i.e. diffusion coefficient rising with the radius (or with a decline
of the density). This case represents the situation when the
gas in the very center is more steady than in the  outer
layers. With our choice of $D_0$ and scaling radius $r_0$ the
predicted profiles are too peaked compared to that observed. 
Of course with such simulations we cannot constrain the behavior
of the diffusion coefficient at even larger radii where the stochastic
motions produced by e.g. mergers may stir and mix the ICM. We can only
conclude that within the core of the cluster diffusion coefficients
declining with radius seem preferable compared to rising ones.

\section{Discussion}
\subsection{Constraints on the velocities and  spatial scales}
\begin{figure}
\plotone{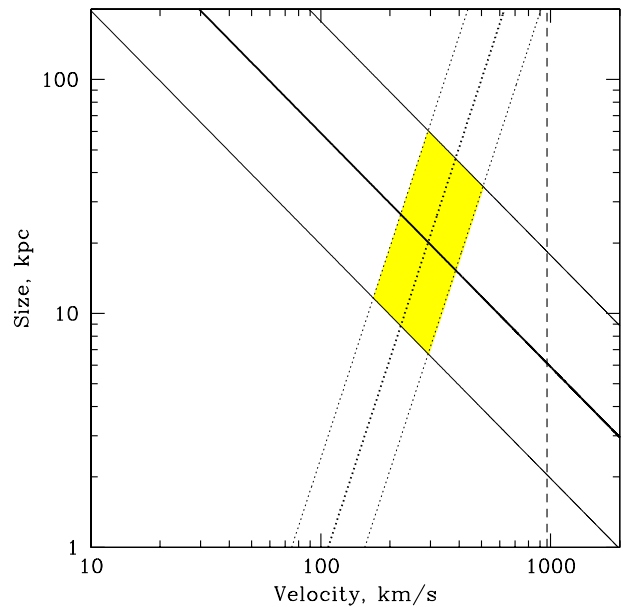}
\caption{Range of the characteristic velocities $v$ 
and spatial scales $l$ of the gas motions  which provide the 
necessary diffusion and
dissipation rates. Along the thick solid line the diffusion
coefficient $D=0.11 v l$ is constant and equal to $2\times10^{29}~{\rm
cm^2~s^{-1}}$. Along the thick dotted line the dissipation/heating rate
$\Gamma_{diss}=0.4~\rho~v^3/l$ is equal to the gas cooling rate
$n^2\Lambda(T)$.  At the intersection of these curves, $v\sim
300~km~s^{-1}$ and $l\sim 20$ kpc, two conditions are satisfied: i)
gas cooling is balanced by the dissipation and ii) the diffusion
coefficient is of the right order. The thin solid and the dotted lines
correspond to a variation of the coefficients in eqns. 8 -\ref{eq:dd}
by factors of 0.3 and 3. The vertical dashed line shows the sound
velocity in the gas.}
\label{fig:vl}
\end{figure}

Summarizing the results of the above section we conclude that the
shape of the observed abundance peak produced by metal ejection from
the central galaxy is consistent with the presence of a diffusive
transport of iron with effective diffusion coefficient of the order of
$D_0\sim 2\times10^{29}~{\rm cm^2~s^{-1}}$. Under the assumption that the
diffusion of iron is due to stochastic gas motions on much smaller
spatial scales (smaller than the radius $r$), one can express the
diffusion coefficient in the form $D\sim C_1 vl$, where $C_1$ is a
dimensionless constant of the order of unity. Thus the value of $D_0$
estimated in the previous section can be considered as a measure of
the product of the characteristic velocity $v$ and the spatial scale
$l$ of the gas motions. On the other hand the gas heating rate due to
the dissipation of the kinetic energy of the same motions also depends
on the combination of $v$ and $l$ and can be written as
$\Gamma_{diss}\sim C_2\rho v^3/l$, where $\rho$ is the gas density and
$C_2$ is a dimensionless constant. Assuming that the disspation of
turbulent motions is the dominant source of heat and that  the
heating rate is equal to the gas cooling rate then
$\Gamma_{diss}\sim\Gamma_{cool}=n^2\Lambda(T)$. The cooling rate can
be easily calculated from the observed gas temperature and density,
thus providing another constraint on $v$ and $l$. Therefore using
$D_0$ and the cooling rate, one can estimate both $v$ and $l$.  The
coefficients $C_1$ and $C_2$ (and the validity of the scaling itself)
in the above expressions depend critically on the character of the gas
motions, which is unknown. One can hope to derive an order of
magnitude estimate using the simplest variants of turbulent flows. For
this estimate we follow the definitions of Dennis \& Chandran (2004)
for $v$ and $l$ which lead to the following expressions for $D$ and
$\Gamma_{diss}$:
\begin{eqnarray}
D&\sim &0.11 v l \\
\Gamma_{diss}&\sim & 0.4 \rho v^3/l. 
\label{eq:dd}
\end{eqnarray}
Using the above definitions and choosing $r=50$ kpc as a
characteristic radius we set $D=2\times10^{29}~{\rm cm^2~s^{-1}}$ and
$\Gamma_{diss}=n^2\Lambda(T)$ to balance cooling and dissipation. One
can plot the resulting constraints in a $v$/$l$ plot (see
Fig.\ref{fig:vl}). The thick solid line shows the combinations of $v$
and $l$ which give the same diffusion coefficient, while along the thick
dotted line the dissipation rate is equal to the cooling rate at $r=50$
kpc. The two pairs of solid and dotted lines show the effect of
varying $C_1$ and $C_2$ by factor of 1/3 and 3 each. The  intersection of the
two curves (bands) gives the locus of the combinations of $l$ and $v$
such that on one hand the diffusion coefficient is approximately equal to
$D_0$ and on the other hand the dissipation rate is approximately equal to
the cooling rate. 
In this figure $l$ and $v$ are the characteristic spatial scales and
velocities of the stochastic gas motions (see Dennis \& Chandran 2004
and references therein). This simplified model suggests that eddies
with size of $\sim$10 kpc and velocities of the order of few hundred
km/s can provide the appropriate iron diffusion and balance gas
cooling losses at the same time. Such velocities and spatial scales
seem reasonable for models where AGN driven outflows are present in
cool cluster cores. In these models the outflow generates motions of
the gas on spatial scales of expanding and rising bubbles (10-15 kpc
for the case of A426) and with velocities characteristic for outflows
(a fraction of the sound speed or few hundred km/s for buoyantly
rising bubbles).  These gas motions eventually dissipate their energy
through viscous heating and/or shocks. The latter mechanism of
dissipation could occur almost without any mixing of metals - if
e.g. there are predominantly radial shock waves, while the former
mechanism provides a closer link between heating and metal
diffusion. While the estimates from Fig.\ref{fig:vl} are very crude,
they are consistent with the assumption that the mixing/dissipation
process plays an important role in clusters with cool cores.  The
presence of differential motions with velocities of the order of few
hundred km/s is also consistent with the lack of spectral signatures
of resonant scattering in the Perseus cluster (Churazov et al. 2004),
although resonant scattering is sensitive to both isotropic (case with
mixing of metals) and radial (no mixing) gas motions. 

We stress again that the uncertainties in the expressions for the
turbulent diffusion coefficient and for the dissipation rate (see
eqns.8-\ref{eq:dd}) are very large and that the numerical results
(Fig.\ref{fig:vl}) should be treated with caution. Moreover the actual
transport of metals in a stratified cluster medium could be very
complicated (e.g. Brueggen 2002), involving on one hand stages of
efficient transport of metals entrained with the rising bubbles and
on the other hand separation of the enriched clumps from the rising
flow due to the entropy contrast. However the above analysis suggests
that cool cores could be a long lived feature (at least few Gyr -
judging from the amount of metals ejected by the BCG - see B\"ohringer
et al. 2004, De Grandi et al. 2004), supported against the
catastrophic cooling by the AGN activity. 

Of course AGN activity is not the only mechanism which can induce
motions in the cluster core. The evidence for gas ``sloshing''
(Markevitch, Vikhlinin \& Forman, 2003) is found in more than 2/3 of
the cooling flow clusters and both mergers or AGN activity may be the
cause. Fujita, Matsumoto \& Wada (2004) argue that the turbulence in
the core is caused by the bulk gas motions of the cluster
gas. Depending on the characteristic spatial and velocity scales of
the motions either direct dissipation or tubulent heat transport may
affect the thermal state of the gas (Dennis \& Chandran, 2004).

\subsection{Alternative scenarios for the abundance peak formation.}
The ejection of metals from the central BCG is not the only
mechanism which could contribute to the formation of  abundance
peaks in cool core clusters. Given the uncertainties in the age of
the system and in the rate of metal ejection by stars, one cannot
exclude other sources of iron based on a simple comparison of the
excess iron mass with the optical luminosity of the central galaxy.
We briefly mention below the most obvious alternative channels of
metal enrichment.

One of the processes which could lead to an enhanced abundance of
metals in the cluster core is gravitational settling (e.g. Fabian \&
Pringle 1977, Rephaeli 1978, Gilfanov \& Sunyaev 1984, Chuzhoy \&
Nusser 2003, Chuzhoy \& Loeb 2004). For clusters with cool cores the
temperature in the core is low (few keV) while the gas density is
high. In such conditions the process of gravitational settling is
rather slow and it is unlikely that it plays an important role in
modifying the observed peaked iron distributions.

More promising seems to be the stripping of metal enriched gas from
smaller subclusters and groups which may have lower entropy than the
bulk of the cluster ICM: group/subcluster gas may sink to the bottom of the
potential well (e.g. Churazov et al. 2003), unless hydrodynamical
instabilities mix it with the ICM before it reaches the cluster
center.

In a similar way, stripping the interstellar medium (ISM) of infalling
spirals or elliptical galaxies (e.g. B\"ohringer et al. 1997,
Toniazzo \& Schindler, 2001) that spend a large part of their time in
orbit at large radii can lead to a concentration of metals in the
cluster center in comparison to the overall galaxy distribution. A
closer inspection of the expected effect of the stripping of the bulk
of the ISM from spirals at column densities of $10^{20} - 10^{21}$
cm$^{-2}$ shows that this should occur at gas densities less than $n_e
\sim 10^{-4}$ cm$^{-3}$ for spiral galaxies falling in face on and at
somewhat larger densities if the galaxies are inclined but little ISM
should reach the inner part of the cluster with gas densities larger
than $n_e \sim 10^{-3}$cm$^{-3}$ (e.g. Cayatte et al. 1994,
B\"ohringer et al. 1997). Thus most of the stripping enrichment of the
ICM should happen outside a radius of 300 kpc and well outside the
region considered here. Stripping should take place both in cooling
flow and non-cooling flow clusters. And unless all non-cooling flow
clusters have experienced a recent merger one would expect abundance
gradients in both types of clusters.

The good correlation of the BCG luminosity and the iron excess mass
(e.g. De Grandi et al. 2004) seems to suggest that the above mentioned
alternative mechanisms do not play a dominant role in the formation of
the central abundance excess.  Nevertheless it is not easy to discard
these mechanisms completely.  One can hope that with the launch of
ASTRO-E it will be possible to derive new constraints on the character
of the gas motions in the cool cores clusters, in particular on the
amplitude and the characteristic spatial scales of the velocity
perturbations in the cluster cores. The anticipated energy resolution
of ASTRO-E spectrometer - FWHM$\sim$ 6-7 eV (see
e.g. \verb#http://heasarc.gsfc.nasa.gov/docs/astroe#) corresponds to
 velocities of $\sim$ 200 km/s for the 6.7 keV line. For comparison
the width of the 6.7 keV line due to thermal motions of iron ions is
$\sim$100 km/s (at a temperature of 3 keV). Therefore it is
realistic to expect the detection of line broadening for 
characteristic stochastic velocities larger than $\sim$100 km/s. The
position of the line can be measured with much higher accuracy and
therefore the presence of large scale motions can be verified. Given
the angular size of a single  spectrometer pixel of $\sim 0.5'$,
the spatial scales one can probe in Perseus are of order of 10
kpc. Thus Perseus observations with ASTRO-E will provide a decisive
test of the models discussed above.

\section{Conclusions}
We show that the heavy metal abundance peak observed in the Perseus
cluster is consistent with a diffusive spreading of metals, provided
that the diffusion coefficient is of the order of $2\times10^{29}~{\rm
cm^2~s^{-1}}$. Stochastic gas motions with characteristic velocities
of the order of few hundred km/s and spatial scales of the order of 10
kpc can provide the diffusive transport of iron and simultaneously a
dissipation rate comparable with the gas cooling rate. Such parameters
seem to be broadly consistent with models where self-regulated AGN
driven outflows are present in the cool cores clusters. ASTRO-E2
observations of Perseus should provide a clear test of the importance
of turbulent mixing in cool cluster cores.

\section*{Acknowledgments}
We would like to thank A.Loeb, A.Fabian and the referee for their valuable
comments. 
P.R. thanks the family of ``La casa delle erbe'', where
this paper was finished, for the use
 of their computer and their nice hospitality.

\label{lastpage}

\addcontentsline{toc}{chapter}{Bibliography}

\begin{thebibliography}{99}




\bibitem[Allen \& Fabian, 1998]{AF98}
Allen, S. W., Fabian, A. C., 1998, MNRAS, 297, L63



\bibitem[Anders \& Grevesse, 1989]{AG89}
Anders E., Grevesse N., 1989, Geochimica et Cosmochimica Acta, 53, 197

\bibitem[\protect\citeauthoryear{Boehringer et 
al.}{1997}]{1997ApJ...485..439B} Boehringer H., Neumann D.~M., Schindler 
S., Huchra J.~P., 1997, ApJ, 485, 439 

\bibitem[B\"ohringer et al. 2001]{Betal01}
B\"ohringer, H., Belsole, E., Kennea, J., et al. 2001, A\&A, 365, L18

\bibitem[B\"ohringer et al. 2004]{Betal04}
B\"ohringer H., Matsushita K., Churazov E., Finoguenov A., Ikebe Y.,
2004, A\&A, 416, 21 

\bibitem[Br\"uggen, 2002 ]{B02}
Br\"uggen, M., 2002, ApJ, 571, L13

\bibitem[Brunzendorf \& Meusinger, 1999 ]{BM99}
Brunzendorf, J., Meusinger H., 1999, ApJS, 139, 141

\bibitem[\protect\citeauthoryear{Cappellaro, Evans, \& 
Turatto}{1999}]{1999A&A...351..459C} Cappellaro E., Evans R., Turatto M., 
1999, A\&A, 351, 459 

\bibitem[\protect\citeauthoryear{Cayatte et 
al.}{1994}]{1994AJ....107.1003C} Cayatte V., Kotanyi C., Balkowski C., van 
Gorkom J.~H., 1994, AJ, 107, 1003 




\bibitem[\protect\citeauthoryear{Cho et al.}{2003}]{2003ApJ...589L..77C} 
Cho J., Lazarian A., Honein A., Knaepen B., Kassinos S., Moin P., 2003, 
ApJ, 589, L77 

\bibitem[\protect\citeauthoryear{Churazov et 
al.}{2002}]{2002MNRAS.332..729C} Churazov E., Sunyaev R., Forman W., B{\" 
o}hringer H., 2002, MNRAS, 332, 729 

\bibitem[\protect\citeauthoryear{Churazov et 
al.}{2001}]{2001ApJ...554..261C} Churazov E., Br{\" u}ggen M., Kaiser 
C.~R., B{\" o}hringer H., Forman W., 2001, ApJ, 554, 261 


\bibitem[Churazov et al. 2003]{Cetal03}
Churazov, E.,Forman, W., Jones, C., B\"oringer, H., 2003, ApJ, 
590, 225

\bibitem[Churazov et al. 2004]{Cetal04}
Churazov, E.,Forman, W., Jones, C., Sunyaev, R., B\"oringer, H. 2004, MNRAS,
347, 29

\bibitem[Chuzhoy \& Loeb, 2004]{CL04}
Chuzhoy, L., Loeb, A., MNRAS, 349, L13

\bibitem[\protect\citeauthoryear{Chuzhoy \& 
Nusser}{2003}]{2003MNRAS.342L...5C} Chuzhoy L., Nusser A., 2003, MNRAS, 
342, L5 

\bibitem[Ciotti et al. 1991]{CDetal91}
Ciotti L.,D'Ercole A., Pellegrini S., Renzini A., 1991, ApJ 376 380


\bibitem[De Grandi et al. 2004]{DELM03}
De Grandi, S.,Ettori, S., Longhetti, M., Molendi, S.,2004, A\&A, 419, 7

\bibitem[De Grandi \& Molendi, 2001]{DM01}
De Grandi S., Molendi S., 2001, ApJ, 551, 153

\bibitem[De Grandi \& Molendi, 2002]{DM02}
De Grandi S., Molendi S., 2002, ApJ 567 163


\bibitem[Dennis \& Chandran, 2005]{DC04}
Dennis, T., Chandran, B., 2005, ApJ, 622, 205




\bibitem[Fabian \&Pringle, 1977]{FP77}
Fabian, A.C., Pringle, J.E., 1977, MNRAS, 181, 5

\bibitem[Fabian, 1994]{F94}
Fabian, A.,C. 1994, ARA\&A, 32, 277

\bibitem[\protect\citeauthoryear{Fabian et al.}{2003}]{2003MNRAS.344L..43F} 
Fabian A.~C., Sanders J.~S., Allen S.~W., Crawford C.~S., Iwasawa K., 
Johnstone R.~M., Schmidt R.~W., Taylor G.~B., 2003, MNRAS, 344, L43 


\bibitem[\protect\citeauthoryear{Finoguenov et 
al.}{2002}]{2002A&A...381...21F} Finoguenov A., Matsushita K., B{\" 
o}hringer H., Ikebe Y., Arnaud M., 2002, A\&A, 381, 21 

\bibitem[\protect\citeauthoryear{Fujita et 
al.}{2004}]{2004ApJ...612L...9F} Fujita Y., Matsumoto T., Wada K.,
2004, ApJ, 612, 9

\bibitem[\protect\citeauthoryear{Fukazawa et 
al.}{1994}]{1994PASJ...46L..55F} Fukazawa Y., Ohashi T., Fabian A.~C., 
Canizares C.~R., Ikebe Y., Makishima K., Mushotzky R.~F., Yamashita K., 
1994, PASJ, 46, L55 

\bibitem[\protect\citeauthoryear{Fukazawa et 
al.}{2000}]{2000MNRAS.313...21F} Fukazawa Y., Makishima K., Tamura T., 
Nakazawa K., Ezawa H., Ikebe Y., Kikuchi K., Ohashi T., 2000, MNRAS, 313, 
21 

\bibitem[Gilfanov \& Sunyaev]{GS84}
Gilfanov, M.R., Sunyaev, R.A.,1984, SvAL, 10, 137



\bibitem[Gastaldello \& Molendi, 2004]{GM04}
Gastaldello, F., Molendi, S., 2004, ApJ, 600, 670

\bibitem[Hernquist 1990]{H90}
Hernquist L., 1990, ApJ, 356, 359




\bibitem[\protect\citeauthoryear{Irwin \& 
Bregman}{2001}]{2001ApJ...546..150I} Irwin J.~A., Bregman J.~N., 2001, ApJ, 
546, 150 


\bibitem[\protect\citeauthoryear{Jones \& 
Forman}{1984}]{1984ApJ...276...38J} Jones C., Forman W., 1984, ApJ, 276, 38 

\bibitem[\protect\citeauthoryear{Kaastra et 
al.}{2004}]{2004A&A...413..415K} Kaastra J.~S., et al. 2004, A\&A, 413, 
415 


\bibitem[Kim \& Narayan 2003 ]{KN03}
Kim, W., Narayan, R. 2003, ApJ, 596L, 139



\bibitem[\protect\citeauthoryear{Markevitch, Vikhlinin, \& 
Forman}{2003}]{2003mecg.conf...37M} Markevitch M., Vikhlinin A., Forman 
W.~R., Matter and Energy in Clusters of Galaxies, ASP Conference Proceedings, Vol. 301. Held 23-27 April 2002 at National Central University, Chung-Li, Taiwan. Edited by Stuart Bowyer and Chorng-Yuan Hwang. San Francisco: Astronomical Society of the Pacific, 2003. ISBN: 1-58381-149-4, p.37


\bibitem[\protect\citeauthoryear{Matsumoto et 
al.}{1996}]{1996PASJ...48..201M} Matsumoto H., Koyama K., Awaki H., Tomida 
H., Tsuru T., Mushotzky R., Hatsukade I., 1996, PASJ, 48, 201 

\bibitem[Matsushita et al. 2002]{Metal02}
Matsushita, K., Belsole, E., Finoguenov, A., B\"ohringer, H., 
2002, A\&A, 386, 77

\bibitem[Matsushita et al. 2003]{Metal03}
Matsushita, K., Finoguenov, A., B\"ohringer, H., 
2003, A\&A, 401, 443

\bibitem[Matsushita et al. 2005]{Metal05}
Matsushita, K., B\"ohringer, H., Takahashi, I., Ikebe, Y., 2005, A\&A,
in press  


\bibitem[Mushotzky \& Loewenstein, 1997 ]{ML97}
Mushotzky, R. F., Loewenstein, M., 1997, ApJ, 481, L63 


\bibitem[\protect\citeauthoryear{Narayan \& 
Medvedev}{2001}]{2001ApJ...562L.129N} Narayan R., Medvedev M.~V., 2001, 
ApJ, 562, L129 

\bibitem[\protect\citeauthoryear{Peterson et 
al.}{2003}]{2003ApJ...590..207P} Peterson J.~R., Kahn S.~M., Paerels 
F.~B.~S., Kaastra J.~S., Tamura T., Bleeker J.~A.~M., Ferrigno C., Jernigan 
J.~G., 2003, ApJ, 590, 207 


\bibitem[Renzini et al. 1993]{Retal93}
Renzini A., Ciotti L., D'Ercole A., Pellegrini S., 1993,ApJ, 419, 52

\bibitem[Rephaeli, 1978]{R78}
Rephaeli, Y., 1978, ApJ, 225, 335

\bibitem[\protect\citeauthoryear{Sanders et 
al.}{2004}]{2004MNRAS.349..952S} Sanders J.~S., Fabian A.~C., Allen S.~W., 
Schmidt R.~W., 2004, MNRAS, 349, 952 




\bibitem[Schmidt et al. 2002]{SFS02}
Schmidt, R.W., Fabian, A.C., Sanders, J.S., 2002, MNRAS, 337, 71


\bibitem[Schombert, 1987]{S87}
Schombert J.\,M., 1987, ApJSS, 64, 643

\bibitem[Schombert, 1988]{S88}
Schombert J.\,M., 1988, ApJ 328 475



\bibitem[\protect\citeauthoryear{Toniazzo \& 
Schindler}{2001}]{2001MNRAS.325..509T} Toniazzo T., Schindler S., 2001, 
MNRAS, 325, 509 

\bibitem[\protect\citeauthoryear{Tozzi et al.}{2003}]{2003ApJ...593..705T} 
Tozzi P., Rosati P., Ettori S., Borgani S., Mainieri V., Norman C. 2003, 
ApJ, 593, 705 


\bibitem[\protect\citeauthoryear{Voigt \& 
Fabian}{2004}]{2004MNRAS.347.1130V} Voigt L.~M., Fabian A.~C. 2004, MNRAS, 
347, 1130 


\end{thebibliography}
\end{document}